\documentclass[preprint,pra,nofootinbib]{revtex4-2} 
\usepackage{amsmath}  % needed for \tfrac, \bmatrix, etc.
\usepackage{amsfonts} % needed for bold Greek, Fraktur, and blackboard bold
\usepackage{graphicx} % needed for figures
\usepackage{color} 
\usepackage{amssymb}
\usepackage{appendix}
\usepackage{enumerate}

\begin{document}

\title{Energy level shift of quantum systems via the scalar electric Aharonov-Bohm effect}

\author{RY Chiao}
\email{raymond\_chiao@yahoo.com}
\affiliation{University of California, Merced, School of Natural Sciences, P.O. Box 2039,
Merced, CA 95344, USA}
\author{H Hart}
\email{hhart3@ucmerced.edu}
\affiliation{University of California, Merced, School of Natural Sciences, P.O. Box 2039,
Merced, CA 95344, USA}
\author{NA Inan}
\email{ninan@ucmerced.edu}
\affiliation{Clovis Community College, 10309 N. Willow, Fresno, CA 93730 USA}
\affiliation{University of California, Merced, School of Natural Sciences, P.O. Box 2039,
Merced, CA 95344, USA}
\affiliation{Department of Physics, California State University Fresno, Fresno, CA 93740-8031, USA}
\author{M Scheibner}	\email{mscheibner@ucmerced.edu}
\affiliation{University of California, Merced, School of Natural Sciences, P.O. Box 2039,
Merced, CA 95344, USA}
\author{J Sharping}
\email{jsharping@ucmerced.edu}
\affiliation{University of California, Merced, School of Natural Sciences, P.O. Box 2039,
Merced, CA 95344, USA}
\author{DA Singleton}
\email{dougs@mail.fresnostate.edu}
\affiliation{Department of Physics, California State University Fresno, Fresno, CA 93740-8031, USA}
\author{ME Tobar}
\email{michael.tobar@uwa.edu.au}
\affiliation{Quantum Technologies and Dark Matter Labs, Department of Physics, University of Western Australia, Crawley, WA 6009, Australia.}

\date{\today}

\begin{abstract}
A novel version of the electric Aharonov-Bohm effect is proposed where the quantum system which picks up the Aharonov-Bohm phase is confined to a Faraday cage with a time varying, spatially uniform scalar potential. The electric and magnetic fields in this region are effectively zero for the entire period of the experiment.  The observable consequence of this version of the electric Aharonov-Bohmn effect is to shift the energy levels of the quantum system rather than shift the fringes of the 2-slit interference pattern. We show a strong mathematical connection between this version of the scalar electric AB effect and the AC Stark effect.   
\end{abstract}

\maketitle
 
\section{The Aharonov-Bohm effect}

The Aharonov-Bohm effect \cite{ab} (AB effect hereafter) shows that in the quantum mechanical context the scalar and vector potentials have a greater reality than is implied by classical E\&M, where the potentials can be eliminated in favor of the fields. The two forms of the AB effect are known as the magnetic or vector AB effect (since this involves the magnetic field coming from the vector potential) and electric or scalar AB effect (since this involves the electric field coming from the scalar potential). The term scalar AB effect is also applied to cases when the system develops an AB phase connected with the scalar interaction ${\bf \mu \cdot B}$ \cite{allman,allman2,wt-lee}. In this work the scalar interaction we focus on is $qV$, {\it i.e.} the scalar coupling between the charge and the electric scalar potential. The original experimental setup of the scalar electric AB effect, which is the focus of this work, is shown in Figure 1. 

The vector AB effect - which occurs by placing an infinite solenoid carrying a constant magnetic flux behind the slits of a 2-slit interference experiment - was first confirmed experimentally by Chambers \cite{chambers} a year after the original theoretical work by Aharonov and Bohm. This initial experimental demonstration of the vector AB effect had some experimental loop-holes, chief among these being that the electrons did not move in a purely ${\bf B}$-field free region. These loop-holes where later plugged in a tour-de-force experiment \cite{tonomura} by Tonomura {\it et al.}, which replaced the unrealistic, infinite solenoid with a micro-scale torus. 

Next, the scalar AB effect connected with the interaction ${\bf \mu \cdot B}$, while not as extensively tested as the magnetic or vector AB effect, has nevertheless received a substantial amount of experimental confirmation - in addition to references \cite{allman,allman2,wt-lee} mentioned above we note the works \cite{shinohara,badurek}. Reference \cite{allman} in particular very clearly lays out the distinction between the scalar AB effect coming from ${\bf \mu \cdot B}$ versus that coming from $q V$. Additionally we will use the term ``scalar electric AB effect" in the present work, since there is a dual formulation of electromagnetism where the electric field can arise from the curl of a dual vector potential, which then leads to a dual AB effect \cite{wei,dowling,tobar}. This leads to a dual ``vector electric AB effect". 

\begin{figure}
\label{fig1}
    \centering
    \includegraphics[scale=0.35]{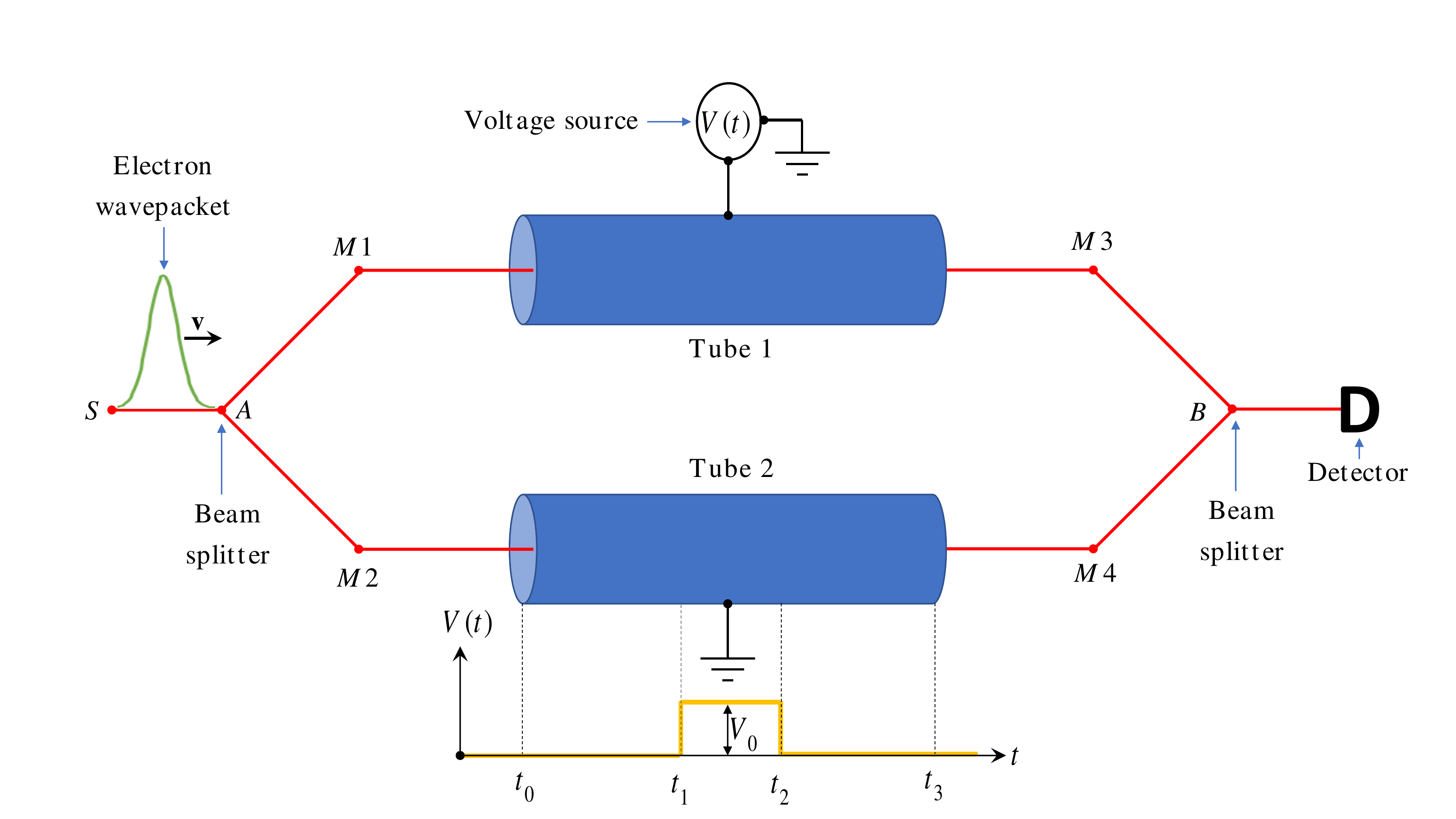}
    \caption{The original proposed set-up for the electric AB effect.}
\end{figure}

Finally, in contrast with the above two versions of the AB effect, the scalar electric AB effect has received much less experimental attention. A prominent experimental verification of the scalar electric AB effect was given in \cite{EAB}. However this experiment measured the effects of both the scalar electric and vector AB effect, rather than the effect of only the scalar electric AB effect. Also at some point in the experiment the electrons did move in a region where the ${\bf E}$-field was non-zero.   

In this work we propose a new version of the scalar electric AB effect, that improves on the set-up in Figure 1. The basic set-up, for our new proposal to test the scalar electric AB effect is given in Figure 2. It consists of a Faraday shell with a time varying voltage on its surface. Outside the Faraday shell there will be both a time varying ${\bf E} (t)$-field and time varying scalar potential $V(t)$. However, inside the Faraday shell the ${\bf E}$-field will be zero; there will be only a time-varying, spatially uniform scalar potential $V(t)$. The quantum mechanical system that we use to register the effect of this $V(t)$ inside the Faraday shell is a gas of hydrogen-like atoms. We propose using rubidium gas since rubidium can more easily be obtained in atomic form, whereas hydrogen generally comes as a molecule, $H_2$. In the next section we discuss some of the basic theoretical aspects of this new version of the scalar electric AB effect. We find is that the time varying, spatial uniform potential, $V(t)$, will split the energy levels into a series of energy levels. In terms of the mathematical analysis this almost identical to the AC Stark effect \cite{autler,DK} where a time varying electric field will split the energy levels of the atom. The difference here is that the energy level splitting occurs in atoms placed in a region with a time varying scalar potential, $V(t)$, but zero electric field ${\bf E} = - \nabla V(t) =0$. This shift in energy levels as a means to probe the AB effect is different from the usual signature of the AB effect, which involves a shift in interference fringes.

\section{Analysis for new proposed scalar electric AB effect}

Here we give the analysis for the new proposed scalar electric AB effect. The quantum system used to probe the scalar electric AB effect is assumed to have a Hamiltonian, $H_0$, for which the solutions to the time-independent Schr{\"o}dinger equation are known {\it i.e.} $H_0 \Psi _i ({\bf x}) = E_i \Psi _i ({\bf x})$. 
To observe the scalar electric AB effect we place the quantum system inside a Faraday shell connected to a time varying voltage as in Figure 2. The new Hamiltonian is  
\begin{equation}
H=H_0+eV(t)  \label{hamiltonian}
\end{equation}%
We take the scalar potential $V(t)$ to be sinusoidal and of the form 
\begin{eqnarray}
\label{Vt}
V(t) &=& 0 ~~~~{\rm for} ~ t < 0 \nonumber \\
V(t) &=& V_0 \cos \Omega t ~~~~{\rm for} ~ t \ge 0
\end{eqnarray}
where $\frac{\Omega}{2 \pi}$ is the frequency and $V_0$ is the amplitude. For $t<0$ where $V(t)=0$ the Hamiltonian is just $H_0$ with wave function solutions $\Psi _i ({\bf x})$, and energy eigenvalues  $E_i$. The time-dependent Schr{\"o}dinger equation for this new system is given by 
\begin{equation}
\label{tdse}
i\hbar \frac{\partial \psi }{\partial t}=H\psi =\left( H_{0}+eV (t) \right) \psi
\end{equation}
Note that while $V(t)$ does depend on time $t$, it does not depend upon $\mathbf{x}$, the position of the electron. In contrast, the
unperturbed Hamiltonian $H_{0}$ will in general depend upon $\mathbf{x}$, but it does not depend on time $t$. Hamiltonians of the form \eqref{tdse} with a piecewise, continuous periodic potential can be solved using Floquet's theorem. We now present a brief summary. 

First we apply the separation-of-variables ansatz
\begin{equation}
\psi (\mathbf{x},t)=X(\mathbf{x})T(t) ~.
\end{equation}
Substituting this ansatz into \eqref{tdse} we find
\begin{equation}
i\hbar \frac{\partial \psi }{\partial t}=i\hbar X\frac{dT}{dt}=\left(
H_{0}+eV\right) XT=TH_{0}X+X\left( eV\right) T
\end{equation}
Dividing this equation by $XT$ and moving the $eV(t)$ term to the left hand side gives
\begin{equation}
-eV+i\hbar \frac{1}{T}\frac{dT}{dt}=\frac{1}{X}H_{0}X
\end{equation}
This equation has the form $f(t)=g\left( \mathbf{x}\right)$
where $f(t)$ is $only$ a function of $t$, and $g\left( \mathbf{x}%
\right) $ is $only$ a function of $\mathbf{x}$. The only way that this can be true is if each function is equal to a constant, $E$ {\it i.e.} $f(t)=g\left( \mathbf{x}\right) =E$. This gives the separated equations
\begin{eqnarray}
-eV+i\hbar \frac{d\ln T}{dt} = E ~~~~
{\text and}~~~~
H_{0}X = EX 
\label{temporal equation}
\end{eqnarray}%
Setting $X=\Psi _{i }({\bf x})$ and $E=E_{i }$, gives the time-independent Schr{\"o}dinger equation
\begin{equation}
H_{0}\Psi _{i }({\bf x})=E_{i }\Psi _{i }({\bf x}) ~,
\end{equation}
which is the known eigenvalue problem for the unperturbed hydrogen-like atom. Integrating the first, temporal equation in \eqref{temporal equation} over $t$ gives,
\begin{equation}
-e\int V(t) dt+i\hbar \int \frac{d\ln T (t)}{dt}dt=\int E_{i }dt \label{int-1}
\end{equation}
Carrying out the integrations in \eqref{int-1} and solving for $T(t)$, gives  
\begin{eqnarray}
T(t) &=&\exp \left( -\frac{i}{\hbar }E_{i }t\right) \exp \left( -\frac{i}{
\hbar }e\int Vdt\right)  \nonumber \\
&=&\exp \left( -\frac{i}{\hbar }E_{i }t-i\alpha \sin \Omega t\right) =\exp
\left( -\frac{i}{\hbar }E_{i }t-i\varphi (t)\right) ~,  \label{T(t) solution}
\end{eqnarray}
where we have defined $\alpha$, the FM depth of modulation parameter 
\begin{equation}
\alpha =\frac{eV_{0}}{\hbar \Omega }
\label{alpha}
\end{equation}

\begin{figure}
\label{fig2}
    \centering
    \includegraphics[scale=0.35]{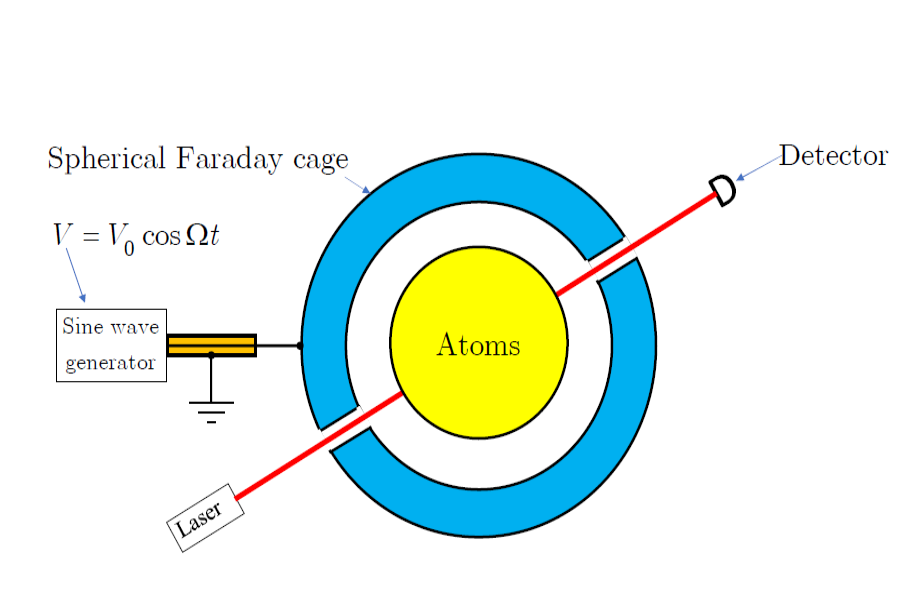}
    \caption{The basic set-up of our proposed test of the scalar electric AB effect}
\end{figure}

Multiplying $X({\bf x}) = \Psi _i ({\bf x})$ and $T(t)$ from \eqref{T(t) solution} gives the wave function, $\psi _i ({\bf r}, t)$, for the full Hamiltonian, $H_0 + e V(t)$ from \eqref{hamiltonian} gives 
\begin{equation}
\psi _{i }(\mathbf{r},t)=\Psi _{i }(\mathbf{r})\exp \left( -\frac{
iE_{i }t}{\hbar }-i\varphi (t)\right) ~. 
\label{psi}
\end{equation}
Note that this new wave function is the original wave function with an added AB phase factor $\exp \left( -i\varphi (t)\right)$. From \eqref{T(t) solution} $\varphi (t)$ is given by 
\begin{equation}
\varphi (t)=\frac{e}{\hbar }\int V(t)dt  = \alpha \sin \Omega t ~, 
\label{phase modulation}
\end{equation}
The expression $\varphi (t)=\frac{e}{\hbar }\int V(t)dt$ above shows that $\varphi (t)$ is the usual scalar electric AB phase. 
Exponentiating the scalar electric AB phase and using the Jacobi-Anger expansion gives
\begin{equation}
\exp \left( - i\varphi (t)\right) =\exp \left( - i\alpha \sin \Omega t\right)
=\sum_{n=-\infty }^{\infty } (-1)^n J_{n}(\alpha )\exp \left( in\Omega t\right)~.
\label{Jacobi-Anger}
\end{equation}
Inserting the result from \eqref{Jacobi-Anger} back into \eqref{psi}, the wave function reads
\begin{eqnarray}
\label{psi-2}
\psi _i(\mathbf{r},t) &=&\Psi _i(\mathbf{r})\sum_{n=-\infty
}^{\infty } (-1)^n J_{n}(\alpha )\exp \left( in\Omega t\right) \exp \left( -\frac{%
iE_i t}{\hbar }\right)  \nonumber \\
&=&\Psi _i(\mathbf{r})\sum_{n=-\infty }^{\infty } (-1)^n J_{n}(\alpha )\exp
\left( -\frac{i\left( E_i-n\hbar \Omega \right) t}{\hbar }\right)
\end{eqnarray}
Thus each energy level $E_i$ will be split into a multiplet $E_i^{(n)}$ with
\begin{equation}
E_i^{(n)}=E_i\pm n\hbar \Omega ,\text{ with}\ n, \text{an integer}
\label{energy-2}
\end{equation}
where $E_i^{(n)}$ are evenly spaced energy levels, with an energy step $\hbar \Omega$. This new energy spectrum, $E_i ^{(n)}$, is of the form of the quasi-energies discussed in \cite{zeldovich}. If one takes the results of equations \eqref{psi-2} and \eqref{energy-2} at face value this would seem to imply a new spectrum with an infinite number of new states labeled by the sideband index $n$. However, from \eqref{psi-2} one finds that the different contributions are weighed by the Bessel functions $J_n (\alpha)$. In Figure 3 we plot $J_n (\alpha)$ for a fixed $\alpha$ as a function of $n$ and find a maximum index given by
\begin{equation}
    \label{nmax}
    n_{max} \approx \alpha ~.
\end{equation}
From Figure 3, where we take $\alpha =1000$, one can see that the Bessel function weighting exponentially drops to zero beyond $n_{max}$. Thus for $n> n_{max}$ the contributions in the sum in \eqref{psi-2} will be suppressed as will the energies $E_i ^{(n)}$ with $n> n_{max}$. Further from Figure 3 one sees that the states and energies which contribute the most and have the largest weighting are those with $n \approx \pm n_{max}$. Taking these $n=n_{max}$ states to dominate one finds that the energy level $E_i$ has been split into two levels: $E_i \pm n_{max} \hbar \Omega$. Recalling that $n_{max} \approx \alpha = \frac{e V_0}{\hbar \Omega}$ then gives new energies levels of  $E_i ^{(n_{max})}= E_i \pm e V_0$.

\begin{figure}
\label{fig3}
    \centering
    \includegraphics[scale=0.35]{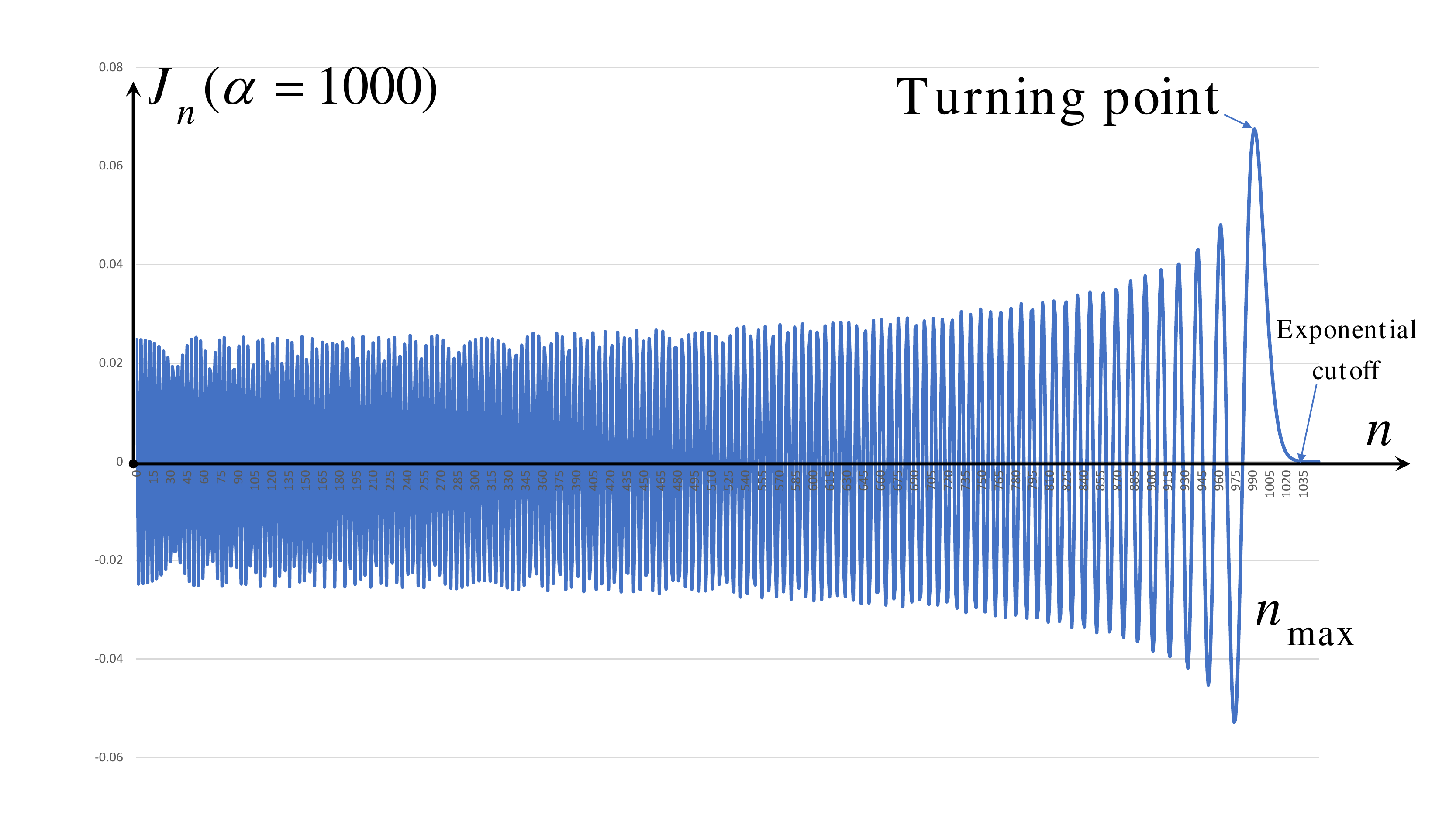}
    \caption{A plot of $J_n (\alpha)$ vs. $n$ for the case when $\alpha =1000$. There are sidebands in the energy $E^{(n)}_i$ which occurs up to some maximum index $n$ given by $n_{max} \approx \alpha$. From the plot one can see that the weighting, $J _n (\alpha)$, is largest when $n=n_{max} \approx \alpha$ and it is this state which contributes the most.}
\end{figure}

This result of the splitting of the energy levels - our version of the scalar electric AB effect - looks {\it almost} mathematically identical to the Autler-Townes effect or AC Stark effect \cite{autler}.  Further in reference \cite{townes} Townes and Schawlow point out that the radio-frequency methodology of frequency modulation (FM) can be applied to quantum mechanical wave functions, whenever their phases undergo sinusoidal frequency modulation due to external perturbations, which mirrors the influence of the external, time varying potential, $V(t)$, on the wave function $\psi_i$ as shown in \eqref{psi-2}. The key difference of course is that for the AC Stark effect the quantum system is in a region of non-zero electric field, while here the quantum system is in a region where the electric field is zero {\it but} the scalar potential is non-zero. 

We now show in detail the mathematical similarity between our version of the scalar electric AB effect and the AC Stark effect. To lay out the similarities we compare our results to those of the AC Stark effect as given in the review article \cite{DK}. Our modified wave function \eqref{psi-2} matches with the wave function from the review article (equation (13) of \cite{DK}). The weighting factor in \cite{DK}, instead of being $(-1)^n J_n (\alpha)$, takes the more complex form 
\begin{equation}
    \label{Cn}
    C_n = \sum _{S=-\infty} ^{+\infty} (-1)^n J_S \left( \frac{\beta E_0^2}{8 \Omega} \right) J_{n+2S} \left( \frac{d E_0}{\Omega}  \right)~,
\end{equation}
where $E_0$ is the magnitude of the electric field that oscillates with frequency $\Omega$, $d$ is a dipole moment, and $\beta$ is a polarizability. In \eqref{Cn} we have changed some of the notation relative to reference \cite{DK} to avoid overlap with our notation. %({\it e.g.} in \cite{DK} $\beta$ is denoted by $\alpha$, but we are already using $\alpha$ for depth of modulation parameter). 
In \cite{DK} there were interactions terms between the electric field and quantum system - one linear and one quadratic in $E_0$. In our case we have only linear terms in the $E_0$ or $V_0$ {\it i.e.} for us $\beta =0$ so that the middle term in the sum in \eqref{Cn} is $J_S (0)$. Now $J_S (0)=0$ {\it except} for $S=0$ ($J_0 (0) =1$). Thus the sum in \eqref{Cn} reduces to only the $S=0$ term namely $C_n = (-1)^n J_n (dE_0/\Omega)$. This exactly matches our weighting $(-1)^n J_n (e V_0/ \hbar \Omega)$. Taking the dipole as $d=e x$ (charge times distance) we find that $dE_0 \sim e x E_0$, noting that a distance times electric field has units of a potential ($x E_0 \sim V_0$), and finally recalling that reference \cite{DK} took $\hbar =1$ one can see that $\frac{d E_0}{\Omega} \sim \frac{e V_0}{\hbar \Omega}$. The $C_n$ weighting in \eqref{Cn} exactly matches our $(-1)^n J_n (\alpha)$ weighting. 

The reason that \cite{DK} had both linear and quadratic electric field terms, whereas we only have linear terms, comes about due to the difference in minimal coupling of the scalar versus vector potential in the time-independent Schr{\"o}dinger equation $i \hbar \frac{\partial}{\partial t} \Psi = \frac{{\hat {\bf p}^2}}{2m} \Psi$. For the vector potential case with a sinusoidal potential ${\bf A} (t) = \frac{E_0}{\Omega} \cos (\Omega t) {\hat {\bf z}}$ minimal coupling is ${\hat {\bf p}}^2 \to ({\hat {\bf p}} - e {\bf A})^2$ which then gives both linear (${\bf A} \sim E_0$) and quadratic (${\bf A}^2 \sim E_0 ^2$) terms. In contrast the minimal coupling of the scalar potential arises via the minimal coupling $i \hbar \frac{\partial}{\partial t} \to i \hbar \frac{\partial}{\partial t}- e V(t)$ with $V(t) = V_0 \cos (\Omega t)$, and this only gives a term linear in the scalar potential. More physically the quadratic term in \cite{DK} is connected with polarization of the quantum system by the electric field, but for our scalar electric AB setup there is no electric field and hence no polarization term. 

This connection between the AC Stark effect and the scalar electric AB effect also supports the earlier statement that only a few of the many energy levels from \eqref{energy-2} are populated and thus play a role. In the AC Stark effect, both theory and experiment \cite{DK} indicate that the $n= \pm n_{max}$ states are dominant, in the limit of low frequency and strong field. It is in this limit that the AC Stark effect and scalar electric AB effect are corrected due to the absence of the quadratic term for the scalar electric AB effect. Thus this close mathematical connection between the AC Stark effect and the scalar electric AB effect indicate that, as for the AC Stark effect, the $n= \pm n_{max}$ states are dominant for the scalar electric AB effect.   

After drawing this exact, mathematical identity between the AC Stark effect and our scalar electric AB effect proposal, we now state that {\it physically} these two are very different. In the case of the AC Stark effect the quantum system is in a region where the electric field is non-zero, so qualitatively (but not quantitatively) all the effects can be explained classically since in classical electromagnetism electric fields can change the energy of the system. In our proposal the scalar potential $V(t)$ is {\it uniform} within the entire interior of the Faraday shell, so that the ${\bf E}$-field is zero inside the Faraday shell and there are \emph{no electrical forces} exerted on the atom. Classically one expects no effect in a Faraday cage {\it i.e.} no energy change for the quantum system. However the connection between the AC Stark effect and our proposal for the scalar electric AB effect, can be compared to Feynman's analysis of the vector AB effect in Volume II of the Feynman Lecture series. In \cite{feynman} Feynman first derives the standard vector AB effect to obtain the shift in the interference pattern when there is an infinite solenoid placed between the slits. He then takes the magnetic field and smears it out into a continuous strip in the region behind the slits and re-analyzes what happens. In this second case the charges going through the slits are subjected to a ${\bf v} \times {\bf B}$ force which changes their momentum and thus shifts the interference pattern. Feynman shows that the shift in the interference pattern is the same in both cases as long as the flux enclosed by the particles path is the same. In the usual AB set-up the shift can only be interpreted as arising from the flux enclosed by the path of the charges, while in the case where the magnetic field is smeared out in a strip behind the slits one can view the shift as coming either from the AB effect and the enclosed flux {\it or} from the shift in the momentum of the charged particles due to a ${\bf v} \times {\bf B}$ force. In an analogous way for the AC Stark effect one can interpret the change in energy levels as coming from either the direct effect of the electric field {\it or} from the influence of the time varying vector potential, ${\bf A} (t)$, giving rise to the electric field. But for our version of the scalar electric AB effect the explanation in terms of the scalar potential, $V(t)$, is the only available explanation for the shift in energy levels.       

We conclude with observations about this version of the scalar electric AB effect:
\begin{itemize}
    \item The scalar electric AB phase, $\varphi (t)$, creates energy level sidebands, equation \eqref{energy-2}, which can be probed via absorption spectroscopy. The dominant energy sidebands occur for $n=n_{max} \approx \alpha$. These energy sidebands are essentially the quasi-energy levels of Zeldovich \cite{zeldovich} and studied later in more detail by Sambe \cite{sambe}. 
    \item The set up here and the analysis leading to wave function \eqref{psi-2} and split energy levels \eqref{energy-2} is mathematically identical to the AC Stark/Autler-Townes effect. In the above we have shown how our proposed set up for the scalar electric AB effect matches the AC stark effect, except for the absence of the quadratic term in the electric field.  
    \item One can question this setup for the scalar electric AB effect since it is possible to gauge away the scalar potential, $V(t) = V_0 \cos (\Omega t)$. A general gauge transformation of the scalar and vector potential is 
    \begin{equation}
        \label{gauge}
        V' = V - \partial_t \lambda ~~~{\rm and}~~~ {\bf A}'={\bf A} + \nabla \lambda~,
    \end{equation}
    with the gauge function $\lambda ({\bf r} , t)$. By choosing $\lambda (t) = \frac{V_0}{\Omega} \sin (\Omega t)$ one can cancel out $V(t) = V_0 \cos (\Omega t)$ so that $V'=0$. Also since $\nabla \lambda =0$ the new vector potential will remain zero, ${\bf A}' =0$. Thus one has gauge transformed away all electromagnetic potentials so how can there be any effect? The resolution to this is that \eqref{gauge} is only half of the gauge transformation. One must also transform the wave function as
    \begin{eqnarray}
    \label{gauge-2}
        \psi ' _i ({\bf r}, t) &=& \exp \left(-i \frac{e}{\hbar} \lambda \right) \psi_i ({\bf r}, t)  = \exp \left( -i \frac{e V_0}{\hbar \Omega} \sin (\Omega t) \right) \Psi _i ({\bf r}) \exp \left( -i \frac{E_i t}{\hbar} \right) \nonumber \\
        &=& \Psi _i ({\bf r} ) \exp \left( -i \frac{E_i t}{\hbar} - i \varphi (t) \right)~.
    \end{eqnarray}
    In arriving at the results in \eqref{gauge-2} we have used results from \eqref{psi} and \eqref{phase modulation}. The wave function $\psi '$ from \eqref{gauge-2}, which is in the gauge where all electromagnetic potentials are zero, matches the wave function $\psi$ from \eqref{psi} which has the non-zero, sinusoidal scalar potential. The results leading  to the energy side bands are gauge invariant. 
    \item This version of the scalar electric AB effect is cleaner than the original proposal shown in Figure 1. We do not have to time the turn-on and turn-off of the potential difference with the charge entering or exiting the metal tubes. Also for the setup in Figure 1 there are always fringing fields, while for the Faraday sphere setup the electric field is zero inside the shell, modulo very small electric and magnetic fields that arise whenever there is a time variation in charges/fields. 
    \end{itemize}
 
\section{Experiment for testing the scalar electric AB effect}

In the preceding section we have shown that our setup for the scalar electric AB effect is mathematically equivalent to the AC Stark effect minus the quadratic term. Thus many experimental tests for the AC Stark effect should apply to our setup for the scalar AB effect, but the quantum systems that act as probes of the effect would be in a region completely free of electric field, in contrast to the AC Stark effect. 

The simplest test for the scalar electric AB effect in our setup is to look for the splitting of the initial energy levels, $E_i$, into two levels $E_i \pm n_{max} \hbar \Omega = E_i \pm e V_0$. This splitting into these two dominant sidebands rather than into a large number of sidebands requires $eV_0 \gg \hbar \Omega$.  For example, one could place hydrogen-like atoms inside the Faraday sphere of Figure 2 with $\Omega= 10^8 ~ s^{-1}$ and $V_0=0.5~m$V so that the two dominant sidebands appear shifted by $\sim 10^{12}~s^{-1} $ in the absorption spectrum. This set of parameters would give $\alpha = \frac{e V_0}{\hbar \Omega} \sim 10^4$ for which $e V_0 \gg \hbar \Omega$ is satisfied. For this type of proposed probe of the scalar AB effect, a vapor of rubidium atoms would be optimal since the expected tunable AB effect feature can be compared with the well-known fine and hyperfine lines.  

Another possible test of our setup of the scalar AB effect is electromagnetically induced
transparency (EIT). Observing  EIT involves two lasers - a probe laser and coupling laser - which are tuned to interact with three quantum states. The probe laser is tuned near resonance between two of the states and measures the absorption spectrum of the transition. A stronger, coupling laser is tuned near resonance at a different transition. By properly selecting the states, the presence of the coupling laser will create a spectral window of transparency which can be detected by the probe laser.
Figure 4 is a sketch of the ``Lambda" version of EIT. This is a three-level scheme to observe EIT. Again we have in mind using rubidium with the three energy levels being the ground state, $5S_{1/2}$, the first excited state, $5P_{1/2}$ and the upper sideband of the ground state, $E_1^{(+)}$. Note that for EIT one needs {\it quantum interference} between two paths in {\it energy} space. The original proposal for both the vector and scalar AB effect from \cite{ab} also worked via interference, but interference in {\it coordinate} space, rather than energy space. Therefore it is possible to use a {\it neutral} atom such as a rubidium atom, in order to observe the scalar AB effect, since one could use the two distinct {\it electronic} pathways, as depicted in Figure 4.

\begin{figure}
\label{fig4}
    \centering
    \includegraphics[scale=0.35]{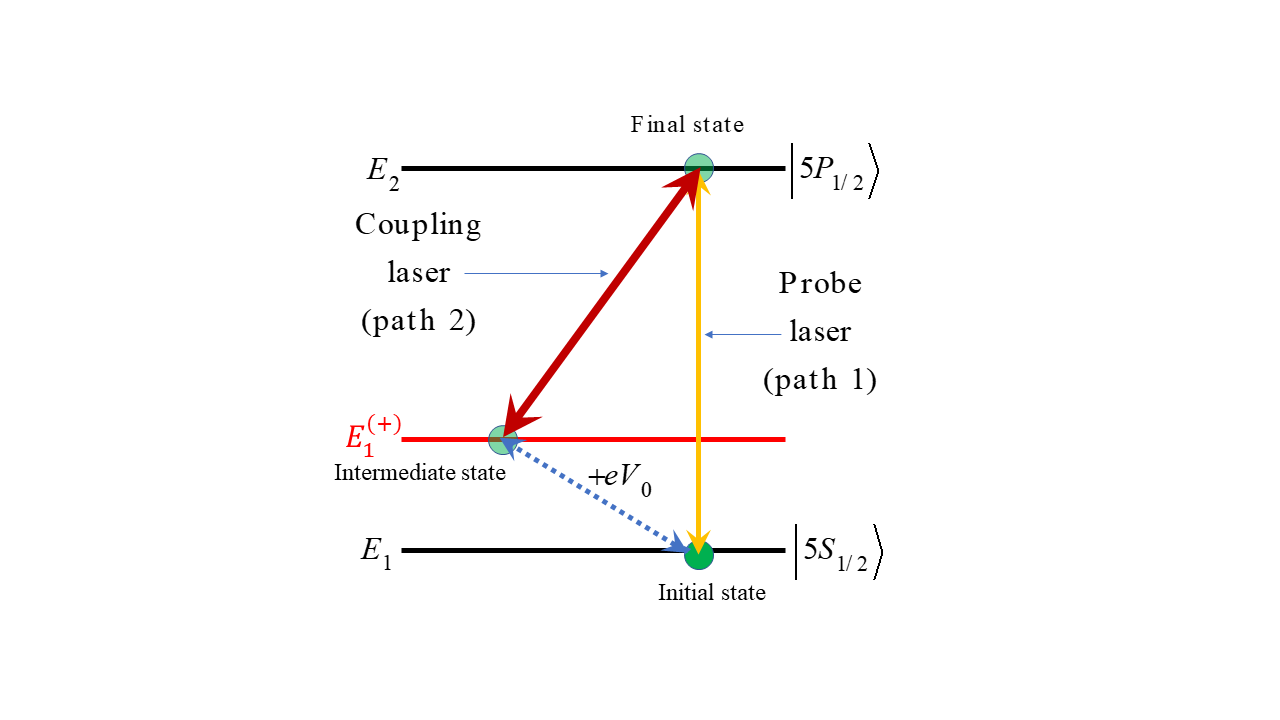}
    \caption{Electromagnetic induced transparency as a demonstration of the scalar AB effect.}
\end{figure}

Two of the three energy levels are the ground state $E_{5S_{1/2}}$ (black) of a two-level atom (rubidium), and its Jacobi-Anger image state $E_{1}^{(+)}$ (red), and the other is the unperturbed excited state $E_{5P_{1/2}}$ (black) of
this atom. In absence of any perturbations, the probe laser (orange) would detect a Lorentzian absorption lineshape in a strong resonant absorption-line transition from $E_{5S_{1/2}}$ to $E_{5P_{1/2}}$.

However, in the presence of the external sine-wave generator driving the outside of the Faraday cage in Figure 2 with an AC voltage $V(t)=V_{0}\cos \Omega t$, there arises the third energy level, $E_{1}^{(+)}$ (red), which
is a Jacobi-Anger image sideband state separated by an energy $+e V_{0}$ from the ground state $E_{5S_{1/2}}$. This image sideband has the same S-wave character as that of the ground state $E_{5S_{1/2}}$. As a result of the action of the coupling laser (deep red), an image of this S-wave state of the ground state is created at the center of the Lorentzian line profile (whose linewidth arises from the $A$ coefficient of spontaneous emission) of the excited P-wave state $E_{5P_{1/2}}$. A selection rule forbids S-wave to S-wave transitions, and this will lead to a central hole of transparency in the middle of the absorption line profile arising from the $E_{5S_{1/2}}$ to $E_{5P_{1/2}}$
transition that is detected by the probe laser, as it is scanned across the strong Lorentzian resonance absorption lineshape of this transition. One can understand this central hole of transparency as arising from the destructive interference between path 1 and path 2 in the energy-level diagram of Figure 4. The interference of the two paths in this energy-level space is analogous to the interference between the two paths in coordinate space in the usual vector AB effect.   

We note that the experimental setup required to observe our version of the scalar electric AB effect is very similar to the setup shown in Figure 1 of \cite{levi}. There is a crucial difference between the setup in reference \cite{levi} and our generic setup from Figure 2. In Figure 2 we have a Faraday shell which screens the quantum system from the electric field, while in reference \cite{levi} the Faraday shell is replaced by a resonating cavity where the quantum system is embedded in a time varying electric field. 

In this section we have focused on using rubidium atoms to probe our version of the scalar electric AB effect. However one could replace rubidium atoms by quantum dots as in \cite{q-dot} where charged quantum dots were used to probe the AC Stark effect.  

\section{Summary and Conclusions}

We have presented a new setup for the scalar electric AB effect which avoids some of the pitfalls of the original set up given in Figure 1 ({\it i.e.} no fringing electric fields, no need to time the turn-on/turn-off of the potential difference with the motion of the charge). In contrast to the vector AB effect where a quantum system develops an AB phase by moving through a time-independent but spatially varying vector potential, ${\bf A} ({\bf r})$, here the quantum system develops an AB phase by sitting at rest in a spatially uniform but time varying scalar potential, $V(t)$. In the vector AB effect with the magnetic flux contained inside an infinite solenoid, the physical consequence of the AB phase is to shift the interference pattern by an amount that depends on the magnetic flux. This shift occurs despite the charged particle moving in a region that is free of magnetic field. 

Similar, to the Feynman analysis of the vector AB effect \cite{feynman} we first have the scalar electric AB effect setup in Figure 2 where the quantum system is in a region with a time varying, but spatially uniform scalar potential, $V(t)$, with no electric field. For this setup we find that the wave function is shifted by an AB phase (see equation \eqref{psi-2}) and the energy spectrum develops sidebands (see equation \eqref{energy-2}). These features of the wave function and energy spectrum are mathematically identical to what is found in the AC Stark or Townes-Autler effect \cite{autler} where the quantum system is placed in a region with a non-zero and time varying electric field. This closes the connection with the Feynman analysis of the vector AB effect since for the AC Stark setup one can use either the electric field or the scalar potential to obtain the effect, whereas for in the setup in Figure 2 one can only see these effects as coming from the scalar electric AB effect, since there is no electric field. There are previous works which have drawn connections between the AC Stark effect and the AB effect \cite{akatsuka,sengstock,imai,numazaki}. However these prior works usually involved the scalar AB effect via the interaction ${\bf \mu \cdot B}$, rather than the scalar electric AB effect via the interaction $qV$, and the experimental signature was still, mostly, a phase shift rather than an energy level shift. 

This opens up the possibility to perform a clean test of the scalar electric AB effect, which looks very different from the original setup of reference \cite{ab} illustrated in Figure 1.

MET is funded by the ARC Centre of Excellence for Engineered Quantum Systems, Grant No. CE170100009 and the ARC Centre of Excellence for Dark Matter Particle Physics, Grant No. CE200100008.


\begin{thebibliography}{99}


\bibitem{ab} Y. Aharonov and D. Bohm, %``Significance of electromagnetic potentials in the quantum theory",
Phys. Rev. {\bf 115}, 484 (1959).

\bibitem{allman} B. E. Allman, {\it et al.},
%A. Cimmino, A. G. Klein, G. I. Opat, H. Kaiser, and S. A. Werner, 
Phys. Rev. A {\bf 48}, 1799 (1993).
%Observation of the scalar Aharonov-Bohm effect by neutron interferometry

\bibitem{allman2} B. E. Allman, W.-T. Lee, O. I. Motrunich, and S. A. Werner,Phys. Rev. A {\bf 60}, 4272 (1999).
%Scalar Aharonov-Bohm effect with longitudinally polarized neutrons

\bibitem{wt-lee}  W.-T. Lee, O. Motrunich, B. E. Allman, and S. A. Werner, Phys. Rev. Lett. {\bf 80}, 3165 (1998).
%Observation of Scalar Aharonov-Bohm Effect with Longitudinally Polarized Neutrons

\bibitem{shinohara} K. Shinohara, T. Aoki, and A. Morinaga,
Phys. Rev. A, {\bf 66}, 042106  (2002).
% Scalar Aharonov-Bohm effect for ultracold atoms

\bibitem{badurek} G. Badurek, {\it et al.},
%H. Weinfurter, R. Gähler, A. Kollmar, S. Wehinger, and A. Zeilinger, 
Phys. Rev. Lett. {\bf 71}, 307 (1993).
%Nondispersive phase of the Aharonov-Bohm effect

\bibitem{chambers} R.G. Chambers, Phys. Rev. Lett. {\bf 5}, 3 (1960).

\bibitem{tonomura} A. Tonomura {\it et al.}, Phys. Rev. Lett. {\bf 56}, 792 (1986).

\bibitem{wei} H. Wei, R. Han, and X. Wei, Phys. Rev. Letts. {\bf 75},  2071, (1995)

\bibitem{dowling} J. P. Dowling, C. P. Williams, and J. D. Franson, Phys. Rev. Letts. {\bf 83}, 2486 (1999). 

\bibitem{tobar} M. E. Tobar, R. Y. Chiao, and M. Goryachev, Sensors {\bf 22}, 7029 (2022).
%Active Electric Dipole Energy Sources: Transduction via Electric Scalar and Vector Potentials

\bibitem{EAB} A. van Oudenaarden, M.H. Devoret, V. Yu. Nazarov, and J.E. Mooij, 
%"Magneto-electric Aharonov–Bohm effect in metal rings". 
{\it Nature}, {\bf 391}, 768 (1998)

\bibitem{autler} S.H. Autler and  C. H. Townes, Phys. Rev., {\bf 100}, 703 (1955).

\bibitem{DK} N.B. Delone and V.P. Krainov, Physics-Uspekhi, {\bf 42}, 669 (1999).

\bibitem{zeldovich} Y. Zel’dovich, Sov. Phys.-JETP {\bf 24}, 1492 (1967).

\bibitem{townes} C.H. Townes and A.L. Schawlow, \emph{Microwave Spectroscopy} (Dover, 1975).

\bibitem{sambe} H. Sambe, Phys. Rev. A {\bf 7}, 2203 (1973).

\bibitem{feynman} R. P. Feynman, {\it The Feynman Lectures on Physics Volume II}, section 15.5 (Basic Books 2011).

\bibitem{levi} F. Levi {\it et al.}, Phys, Rev. A {\bf 93}, 023433 (2016).

\bibitem{q-dot} T. A. Wilkinson, {\it et al.}, Appl. Phys. Lett. {\bf 114}, 133104 (2019).

\bibitem{akatsuka} T. Akatsuka, {\it et al.}, 
%K. Tanihara, K. Komito, K. Ooi, and A. Morinaga,
Phys. Rev. A {\bf 86}, 023418 (2012).
%Dispersion-shaped ac Stark phase shift of Ca intercombination transitions with a time-domain atom interferometer

\bibitem{sengstock} K. Sengstock, {\it et al.},
%U.Sterr, G.Hennig, D.Bettermann, J.H.M{\"u}ller, and W.Ertmer,
Opts. Comm.,  {\bf 103}, 73 (1993).
%Optical Ramsey interferences on laser cooled and trapped atoms, detected by electron shelving

\bibitem{imai}  H. Imai, Y. Otsubo, and A. Morinaga, Phys. Rev. A {\bf 76}, 012116 (2007).
%Evaluation of the geometric phase of a two-level atom manipulated on the Bloch sphere using a time-domain atom interferometer

\bibitem{numazaki} K. Numazaki, H. Imai, and A. Morinaga, Phys. Rev. A {\bf 81}, 032124 (2010).
%Measurement of the second-order Zeeman effect on the sodium clock transition in the weak-magnetic-field region using the scalar Aharonov-Bohm phase



\end{thebibliography}
\end{document}